\documentclass{article}
\usepackage{spconf,amsmath,graphicx,hyperref}

\usepackage{xcolor,soul,framed} 

\colorlet{shadecolor}{yellow}
\usepackage{array}
\usepackage{url}

\usepackage{cite}
\usepackage{booktabs}
\usepackage{balance} 
\usepackage{multirow}
\usepackage{multicol}
\usepackage{amsfonts}
\usepackage{float}
\usepackage{bbold}

\usepackage{hyperref}
\usepackage{etoolbox,siunitx}
\robustify\bfseries
\usepackage{balance}

\usepackage{amssymb}
\usepackage{pifont}
\newcommand{\cmark}{\ding{51}}%
\newcommand{\xmark}{\ding{55}}%

\usepackage{scalerel}

\title{LuSeeL: Language-queried Binaural Universal Sound Event Extraction and Localization}
\name{Zexu Pan, Shengkui Zhao, Yukun Ma, Haoxu Wang, Yiheng Jiang, Biao Tian, Bin Ma}
\address{
  Tongyi Lab, Alibaba Group, Singapore \\
  }
\begin{document}
\ninept
\maketitle
\setlength{\abovedisplayskip}{4pt}
\setlength{\belowdisplayskip}{4pt}

\begin{abstract}
Most universal sound extraction algorithms focus on isolating a target sound event from single-channel audio mixtures. However, the real world is three-dimensional, and binaural audio, which mimics human hearing, can capture richer spatial information, including sound source location. This spatial context is crucial for understanding and modeling complex auditory scenes, as it inherently informs sound detection and extraction. In this work, we propose a language-driven universal sound extraction network that isolates text-described sound events from binaural mixtures by effectively leveraging the spatial cues present in binaural signals. Additionally, we jointly predict the direction of arrival (DoA) of the target sound using spatial features from the extraction network. This dual-task approach exploits complementary location information to improve extraction performance while enabling accurate DoA estimation. Experimental results on the in-the-wild AudioCaps dataset show that our proposed \textbf{LuSeeL} model significantly outperforms single-channel and uni-task baselines.

\end{abstract}

\begin{figure*}[t]
\centering
\includegraphics[width=\linewidth]{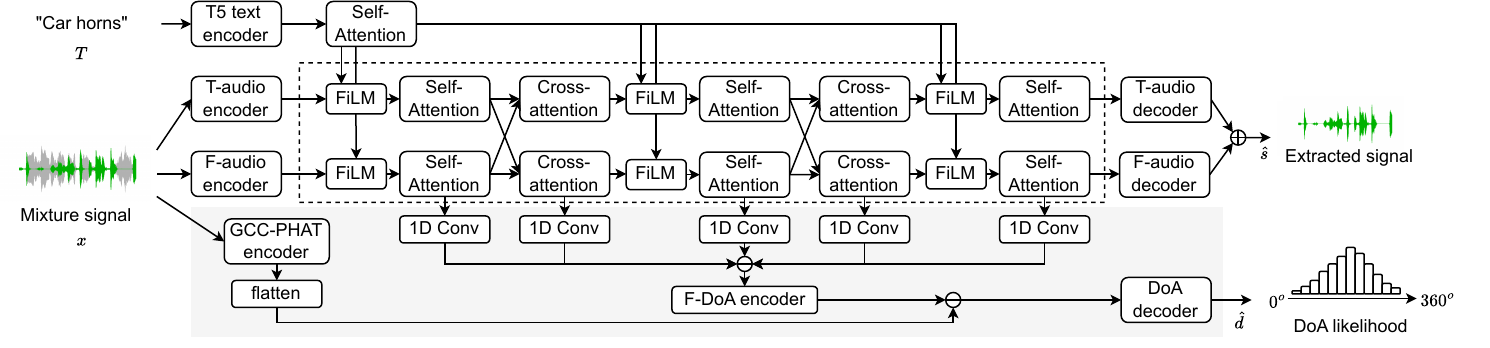}
\vspace{-4mm}
\caption{Our proposed \textbf{LuSeeL}, which jointly performs binaural Universal Sound Event Extraction and Localization using language descriptions.  The symbols $\otimes$ and $\ominus$ represent element-wise multiplication and channel-wise concatenation, respectively.
}
\label{fig:network}
\end{figure*}

\section{Introduction}

Audio scene analysis has emerged as a critical research area in human-machine interaction~\cite{gong2023joint,owens2018audio}, driven by its potential to advance applications in augmented reality, robotics, and assistive technologies. The ability to detect~\cite{li2022atst}, extract~\cite{liu2023separate}, and localize~\cite{zhao2024text} ambient sounds is fundamental for enabling systems to perceive and respond to complex auditory environments. In robotics, spatial sound understanding enables autonomous agents to navigate and interact with their surroundings. Similarly, in augmented reality, accurate sound localization enhances user immersion by synchronizing auditory cues with visual content. Furthermore, these capabilities are essential for processing raw audio from large-scale datasets, which is vital for training modern audio generation models. Therefore, there is an increasing demand for robust universal sound extraction and localization algorithms.

Universal sound event extraction algorithms have gained significant attention in recent years~\cite{zhao2024universal,wisdom2020unsupervised,ma2024clapsep,kong2023uss,fuss2021scott}. The primary challenge lies in detecting and isolating a target sound event from diverse and overlapping acoustic environments, as well as separating it from spectral interference caused by concurrent sound sources. To extract universal sound events, recent approaches have explored cross-modal techniques, including using visual prompts (e.g., ClipSep~\cite{dong2023clipsep}), text-based cues (e.g., AudioSep~\cite{liu2023separate}), and audio-based guidance (e.g., UniSep~\cite{wang2025unisep}). OmniSep~\cite{cheng2025omnisep} further advances this paradigm by integrating visual, textual, and auditory cues into a unified framework.

Despite the advancements, many existing universal sound extraction methods face limitations in handling overlapping sounds, as their performance often relies solely on external prompt cues. This dependency is particularly problematic in complex acoustic conditions, where multiple sound sources coexist. Especially, their use of single-channel audio restricts the ability to capture spatial information, which is critical for accurately understanding the relationships between different sound sources. In contrast, binaural audio, which mimics human hearing through dual-channel recordings, provides rich spatial cues that enhance the accuracy of sound event detection and localization. These cues serve as an additional, implicit reference for isolating the target sound. In practice, applications such as augmented reality, robotics, and assistive technologies heavily depend on spatial awareness to function effectively.

On the other hand, sound event localization and detection have also gained significant attention in recent years~\cite{adavanne2018sound,niu2023experimental,politis2020dataset,jiang2024exploring}, particularly following the launch of the Detection and Classification of Acoustic Scenes and Events (DCASE) Task 3~\cite{politis2020overview}. Most existing works focus on audio-only methods to predict a limited set of predefined classes. Recently, text-queried target sound event localization~\cite{zhao2024text} has advanced the field by enabling universal sound event localization through text-based queries.

We advocate that sound event localization and separation are inherently interdependent tasks, where localization capabilities can enhance the accuracy of sound event extraction. Therefore, in this work, we move beyond single-channel universal sound event extraction and propose a multi-modal framework that jointly extracts and localizes sound events from binaural audio. This framework leverages free-form text descriptions, i.e., a keyword, a set of keywords, or a caption, as a prompt, since text provides greater flexibility and natural alignment with human communication. Such an approach is well-suited for applications like virtual reality and human-robot interaction, where adaptability and intuitive user input are critical.

We propose LuSeeL, a Language-queried binaural Universal Sound Event Extraction and Localization model. It employs a dual-domain hybrid transformer architecture (HTDemucs~\cite{rouard2022hybrid}) as the extraction backbone, which jointly processes the audio signal in both time and spectral domains. A T5~\cite{t52020} text encoder is used to encode text prompts into conditional cues for the extraction process. These text embeddings are fused with the audio features via FiLM~\cite{Zhao_2024_ICML} layers into both the time and spectral representations. The spectral representation, together with generalized cross-correlation with phase alignment (GCC-PHAT) features, is then used to predict the direction of arrival (DoA) of the target sound.
Experimental results on the in-the-wild AudioCaps dataset demonstrate that, on both 2 and 3-source mixtures, LuSeeL significantly outperforms the single-channel extraction baseline, showing the importance of spatial features in separating sources at different angular locations. LuSeeL also significantly outperforms uni-task extraction or localization baselines, showing the effectiveness of the extracted spectral representation in DoA estimation, as well as the role of the DoA task in inherently enhancing spatial awareness during extraction.

\section{LuSeeL}

Our proposed model is illustrated in Fig.~\ref{fig:network}. Given a binaural mixture signal $x$, composed of the target sound event $s$ and interfering sources $b$, the model aims to perform two tasks jointly: (1) extracting the target signal $\hat{s}$ to approximate $s$, and (2) predicting the target signal's DoA likelihood $\hat{d}$ (in degrees) to approximate the ground truth azimuth $d$, conditioned on a text description $T$ such as ``car horns".

\subsection{Model architecture}

\subsubsection{Text encoder module}
We employ a T5 text encoder~\cite{t52020} with frozen weights to encode the text prompts, preserving its pre-trained capacity for generating rich semantic text representations. To better align the text features with the audio modality, we pass the T5 embeddings through additional trainable self-attention layers, adaptively projecting them into a shared embedding space suitable for cross-modal fusion.

\subsubsection{Audio encoder and decoder modules}
We design the audio encoder and decoder following~\cite{rouard2022hybrid}, employing a dual-domain architecture that processes the mixture signal in both the time and spectral domains. Specifically, there is a time-domain audio encoder (T-audio encoder) and a frequency-domain audio encoder based on the Short-Time Fourier Transform (F-audio encoder), which extract time-domain and spectral representations, respectively. Each encoder consists of four convolutional layers that transform the input waveform into frame-level embeddings. 

Similarly, the T-audio and F-audio decoders each comprise four deconvolutional layers to reconstruct the extracted signals back to the waveform domain. The final output $\hat{s}$ is obtained by element-wise addition of the two decoded streams, effectively combining complementary temporal and spectral information.

\subsubsection{Signal extractor module}
The signal extractor module is enclosed in the dotted box in Fig.~\ref{fig:network}. It extracts target audio embeddings through parallel time-domain and spectral-domain processing paths, with guidance from the text embedding from the text encoder and self-attention output.
Each path employs a dedicated stack of self-attention layers to model temporal or spectral-temporal dependencies. Cross-attention layers are inserted between self-attention layers across the two streams to facilitate the exchange of complementary information. The text embedding is fused into both the time and spectral streams before each self-attention layer using Feature-wise Linear Modulation (FiLM), enabling conditional feature adaptation based on the linguistic context.

\subsubsection{Localization module}
The localization module is shown in the bottom portion of Fig.~\ref{fig:network}, highlighted with a gray background. It comprises four main components: a GCC-PHAT encoder, a series of 1D convolutional (1D Conv) layers, an F-DoA encoder, and a DoA decoder.

The GCC-PHAT encoder computes GCC-PHAT features from the binaural mixture signal, capturing inter-channel time differences useful for direction estimation. From the signal extractor, the outputs of each self-attention and cross-attention layer in the spectral stream are first passed through individual 1D Conv layers to extract localization-specific spectral features. These features are then concatenated along the channel dimension and fed into the F-DoA encoder, which consists of several 1D convolutional layers designed to progressively reduce the time dimension while preserving spatial discriminability. The resulting compressed representation is concatenated with the flattened GCC-PHAT features along the channel axis and subsequently processed by the DoA decoder, composed of several fully connected (linear) layers followed by a sigmoid activation to produce a likelihood distribution over azimuth angles.

\subsection{Objective functions}
We adopt the hybrid time-frequency domain loss $\mathcal{L}_{\text{signal}}(\hat{s}, s)$~\cite{pan2022hybrid} as the objective function for training the audio extraction task, which combines two complementary components: the negative scale-invariant signal-to-noise ratio (SI-SNR) loss~\cite{le2019sdr}, denoted as $\mathcal{L}_{\text{SI-SNR}}$, and a frequency-domain multi-resolution delta spectrum loss, $\mathcal{L}_{\text{freq}}$. The $\mathcal{L}_{\text{SI-SNR}}$ optimizes the similarity between the extracted and ground-truth target signals in a scale-invariant manner, focusing on waveform fidelity. In parallel, $\mathcal{L}_{\text{freq}}$ enforces consistency in spectral dynamics across multiple time-frequency resolutions, capturing fine-grained temporal changes in the spectrogram. Notably, $\mathcal{L}_{\text{freq}}$ is scale-variant, which encourages the model to preserve the absolute energy levels of the target source, ensuring that the enhanced signal reflects its true relative intensity in the mixture: 
\begin{equation}
\mathcal{L}_{\text{signal}}(\hat{s}, s)=\mathcal{L}_{\text{SI-SNR}}(\hat{s}, s)+\mathcal{L}_{\text{freq}}(\hat{s}, s)
\end{equation}

We adopt the Mean Squared Error (MSE) as the objective function for training the localization task~\cite{qian2021multi}. The DoA is discretized into $N=360$ azimuth bins, and the ground-truth DoA $d$ is represented as a Gaussian-smoothed label over these bins to account for angular uncertainty. Specifically, the target distribution is defined as:
\[
y(\theta_i) = \mathcal{N}(\theta_i; d, \sigma^2),
\]
where $\theta_i$ denotes the center of the $i$-th angular bin, $d$ is the true azimuth angle, and $\sigma^2=5$ controls the smoothing radius. The predicted DoA likelihood $\hat{d} \in \mathbb{R}^N$ is a probability-like distribution over the same $N$ bins. The MSE loss is computed between $\hat{d}$ and the soft target $y$:
\begin{equation}
\mathcal{L}_{\text{MSE}} = \sum_{i=1}^{N} \left( \hat{d}_i - y(\theta_i) \right)^2 = \|\hat{d} - y\|^2.
\end{equation}

The overall training objective combines the audio reconstruction and localization losses in a weighted manner. Specifically, we optimize:
\begin{equation}
\mathcal{L}_{\text{total}} = \mathcal{L}_{\text{signal}} + \gamma \cdot \mathcal{L}_{\text{MSE}},
\end{equation}
where $\gamma$ is a balancing coefficient set to 10 in our experiments. Gradients from the localization loss $\mathcal{L}_{\text{MSE}}$ are allowed to propagate back through the audio extraction network, enabling end-to-end optimization of shared features. This facilitates the learning of a joint audio-text-spatial representation that supports both high-fidelity audio extraction and accurate spatial localization. Furthermore, we hypothesize that directional information acts as a complementary cue that can enhance audio extraction performance by providing additional contextual constraints on the target source's location.

\section{Experimental setup}

\subsection{Dataset}
Following~\cite{zhao2024text}, we use the AudioCaps dataset~\cite{audiocaps} to evaluate our proposed method with baselines. AudioCaps is a large-scale dataset containing approximately 46K audio clips paired with human-written textual descriptions, collected via crowdsourcing on the AudioSet corpus, and has been widely adopted in audio source separation research~\cite{liu2023separate}. 

To maximize data diversity and enhance model generalization, we dynamically simulate binaural 2-source and 3-source mixtures during training. For each mixture, one clip is selected as the anchor (target). The interfering clips are normalized to match the anchor's energy level and assigned a random signal-to-noise between $-5$~dB and $+5$~dB. Each source is also assigned a random azimuth angle uniformly sampled from $0^\circ$ to $360^\circ$. We then convolve each signal with a head-related transfer function to simulate spatialized binaural audio, and sum individual signals together to form the final mixture. All audio signals are sampled at 16 kHz, with a clip length of 10 seconds.

\subsection{Baselines}
We compare our LuSeeL with several baseline methods:

\par
\noindent
\textbf{T-HTDemucs}~\cite{rouard2022hybrid} is a language-queried, single-channel universal sound event extraction network. Like LuSeeL, it employs a hybrid Transformer-Demucs architecture and is conditioned on a T5 text encoder. However, it is a single-task model designed solely for audio extraction, without any localization capability.

\par
\noindent
\textbf{MLP-GCC}~\cite{qian2021multi, zhao2024text} is a language-queried binaural sound event localization network. It employs a T5 text encoder for conditioning and includes a GCC-PHAT encoder to capture interaural cues. The text and GCC-PHAT features are concatenated and directly fed to the DoA decoder, without leveraging spectral representations from a dedicated audio extraction pathway. This model performs only localization.

\par
\noindent
\textbf{LuSeeL$^{\dagger}$} is an ablated variant of our full model, configured as a language-queried binaural universal sound event extraction network. It retains the signal extraction pathway of LuSeeL but removes the localization module (highlighted in gray in Fig.~\ref{fig:network}), making it a single-task system focused exclusively on extraction.

\par
\noindent
\textbf{LuSeeL$^{\circ}$} is an ablated variant of our full model, which removes the GCC-PHAT encoder. Therefore, localization relies solely on spatial embeddings derived from the sound extraction network. It performs both audio extraction and localization.

\subsection{Network configurations}
The self-attention layers after the T5 text encoder consist of 5 transformer encoder layers with an embedding dimension of 512, 2 attention heads, and a feed-forward hidden size of 1024. For the audio encoders and decoders, and the self- and cross-attention blocks in the signal extractor, we adopt the architectural settings from~\cite{rouard2022hybrid}. Specifically, both the time-domain and spectral-domain processing paths employ 3 self-attention blocks and 2 cross-attention blocks each.

The 1D Conv layers that project audio spectral features for localization have an input dimension of 1256, an output dimension of 100, and a kernel size of 1. The F-DoA encoder comprises three 1D Conv layers with configurations: (input: 500, output: 512, kernel: 1), (512, 256, 1), and (256, 1, 1). ReLU activation and dropout with a rate of 0.1 are applied between each 1D Conv layer.
The DoA decoder consists of six linear layers with input-output dimensions: (888, 1024), (1024, 1024), (1024, 1024), (1024, 1024), (1024, 1024), and (1024, 360). Batch normalization, ReLU activation, and dropout with a rate of 0.1 are applied between each linear layer.

\subsection{Optimization}
All experiments in this work follow the same optimization setup. Training is conducted using distributed data-parallel processing across four Tesla 80 GB A800 GPUs, resulting in an effective batch size of 128. We employ the AdamW optimizer with an initial learning rate of $1 \times 10^{-4}$ and apply a linear learning rate warm-up over the first 5,000 steps. The learning rate is halved whenever the best validation loss does not improve for six consecutive epochs, and training is terminated early if no improvement is observed for ten consecutive epochs.

\begin{table*}
    \centering
    \sisetup{
    detect-weight, 
    mode=text, 
    tight-spacing=true,
    round-mode=places,
    round-precision=1,
    table-format=2.2
    }
    \caption{Results for target audio extraction and localization on both the 2-source and 3-source mixture datasets. For audio extraction, SI-SNRi and SDRi are measured in dB. For localization, the direction recognition accuracy within a ±5$^\circ$ collar (in \%) and the MAE in degrees are presented. Each model is assigned a unique system number (Sys. \#). The table summarizes various models differing in task type, use of GCC-PHAT features in the localization module, and the number of audio channels used.} 
    \begin{tabular}{cccccc |SS|SS}
       \toprule
        \multirow{2}*{Sys. \#}   &\multirow{2}*{Sources}    &\multirow{2}*{Model}   &\multirow{2}*{Tasks} &\multirow{2}*{GCC-PHAT}   &\multirow{2}*{Channels} &\multicolumn{2}{c|}{Extraction}&\multicolumn{2}{c}{Localization}\\ 
        &&&&& &{SI-SNRi} &{SDRi}  &{Accuracy} &{MAE}\\ 
        \midrule
         1   &\multirow{5}*{2}   &T-HTDemucs~\cite{rouard2022hybrid}         &Extraction&\xmark  &1     &7.7097     &8.6932    &{-}    &{-}\\
         2    &  &MLP-GCC~\cite{qian2021multi, zhao2024text} &Localization&\cmark   &2                  &{-}    &{-}    &41.147    &51.645\\
         3    &  &LuSeeL$^{\dagger}$ &Extraction&\xmark       &2                                        &17.5652     &18.799    &{-}    &{-}\\
         4    &  &LuSeeL$^{\circ}$   &Both&\xmark             &2                                        &20.0738    &21.3685    &88.029    &7.1802\\
         5    &  &LuSeeL             &Both&\cmark             &2                                        &20.2682     &21.6057    &89.940   &7.0018\\
        \midrule
         6  &\multirow{5}*{3}     &T-HTDemucs~\cite{rouard2022hybrid}         &Extraction&\xmark  &1    &6.8375     &7.70673    &{-}    &{-}\\
         7    &  &MLP-GCC~\cite{qian2021multi, zhao2024text} &Localization&\cmark     &2                &{-}    &{-}    &26.490    &65.67\\
         8    &  &LuSeeL$^{\dagger}$ &Extraction    &\xmark         &2                                  &10.495     &11.253    &{-}    &{-}\\
         9    &  &LuSeeL$^{\circ}$   &Both          &\xmark         &2                                  &13.3442    &14.157     &75.05 &17.150\\
         10    &  &LuSeeL             &Both          &\cmark         &2                                  &12.860     &13.6986    &77.83    &15.457\\
        \bottomrule
    \end{tabular}
    \label{tab:results}
\end{table*}

\section{Results}
For audio extraction, we report the improvement in Signal-to-Noise Ratio (SI-SNRi) and the improvement in Signal-to-Distortion Ratio (SDRi), computed as the difference between the extracted output and the unprocessed mixture with respect to the ground truth signal. For target audio localization, we evaluate two metrics: the accuracy of predicting the true direction within a tolerance collar of ±5$^\circ$, and the mean absolute error (MAE) in degrees. The lower the better for MAE, and the higher the better for all other metrics. All models are trained on both the 2-source and 3-source mixture datasets, and results are reported separately for each scenario.

\begin{figure*}[t]
\begin{minipage}[t]{.33\linewidth}
  \centering
  \centerline{\includegraphics[width=\linewidth]{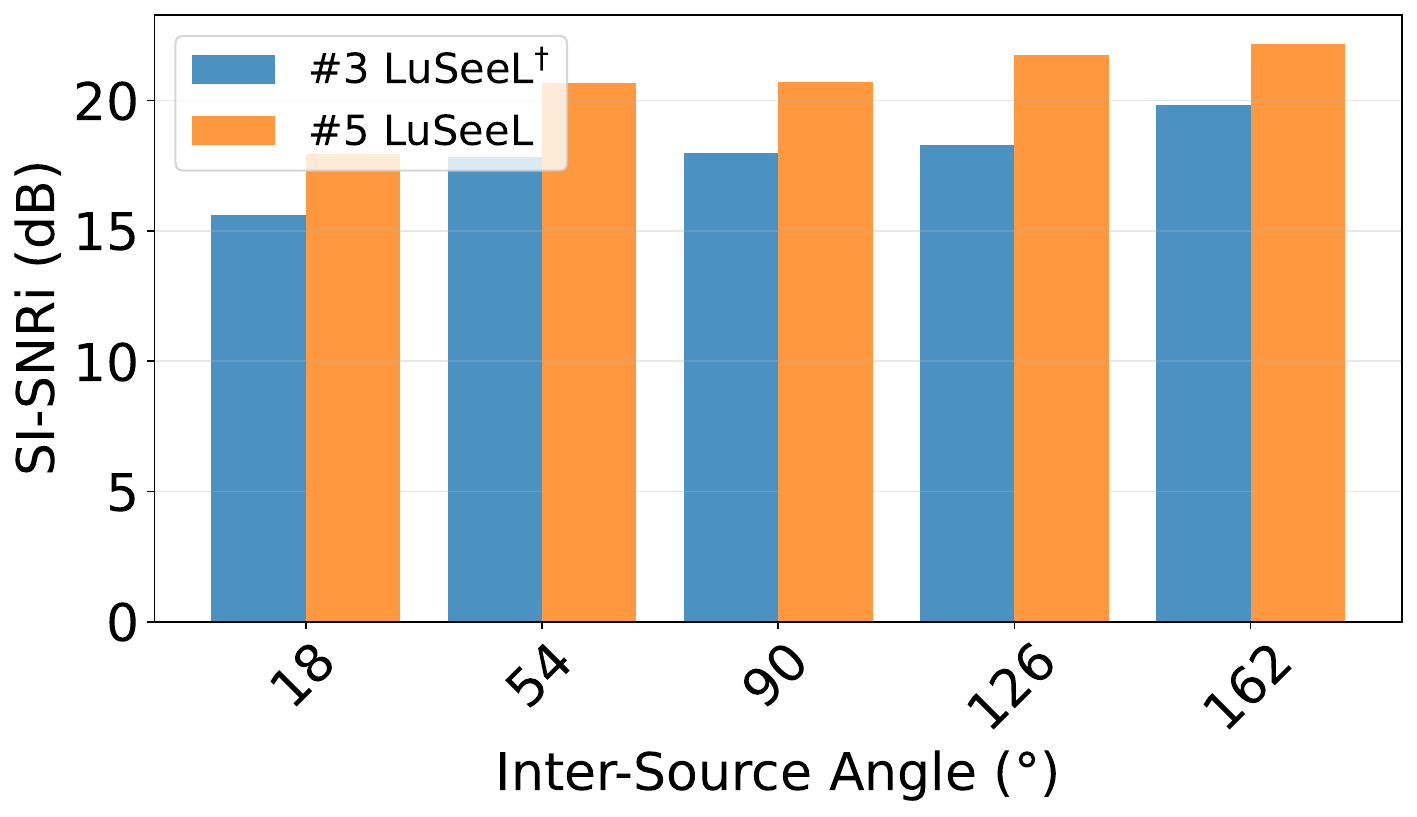}}
  \vspace*{-2mm}
  \caption{The 2-source SI-SNRi histogram of various inter-source separation angles.}\medskip
  \label{fig:histogram_sisnr}
\end{minipage}
\hfill
\begin{minipage}[t]{.33\linewidth}
  \centering
  \centerline{\includegraphics[width=\linewidth]{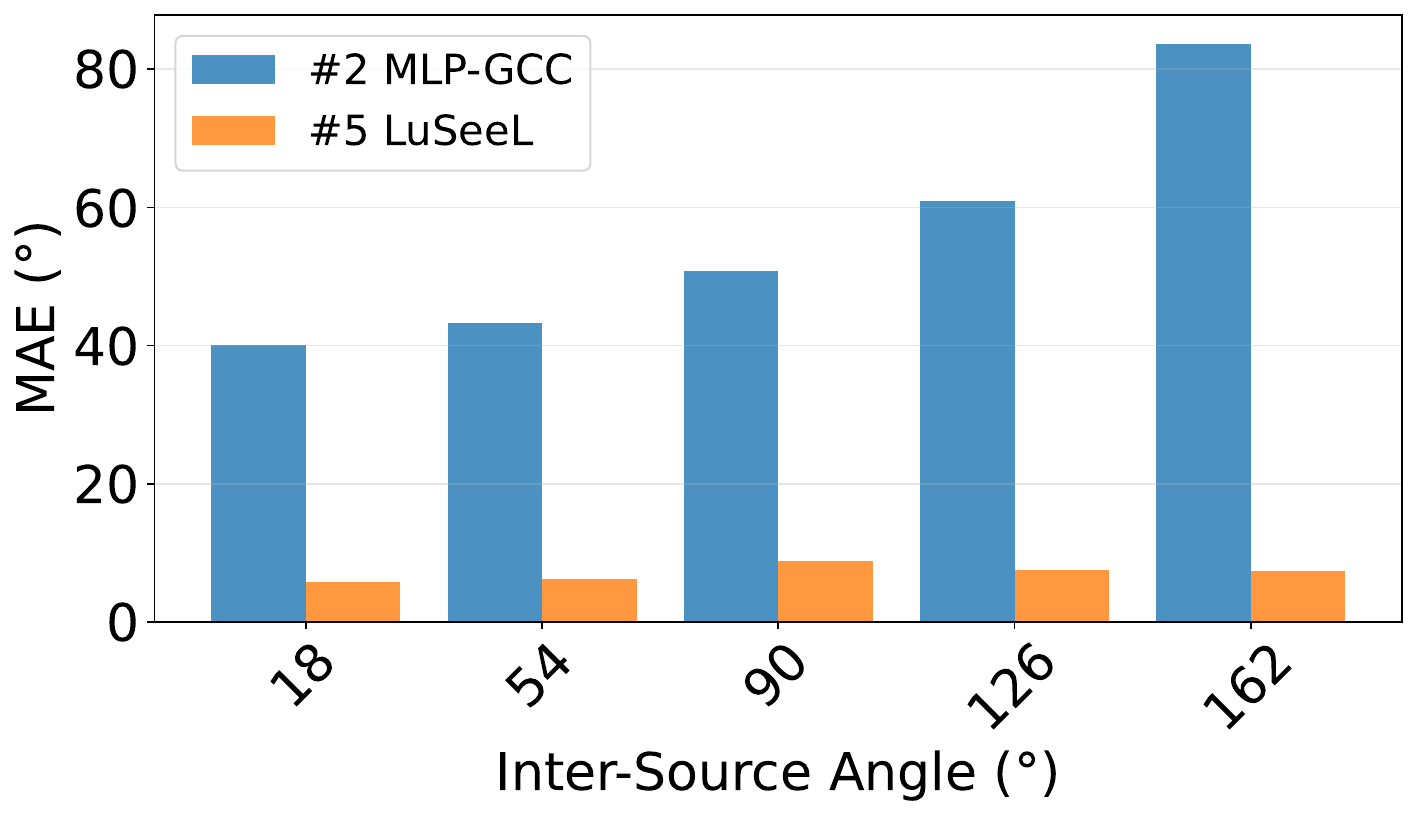}}
  \vspace*{-2mm}
  \caption{The 2-source MAE histogram of various inter-source separation angles.}\medskip
  \label{fig:histogram_mae}
\end{minipage}
\hfill
\begin{minipage}[t]{.33\linewidth}
  \centering
  \centerline{\includegraphics[width=\linewidth]{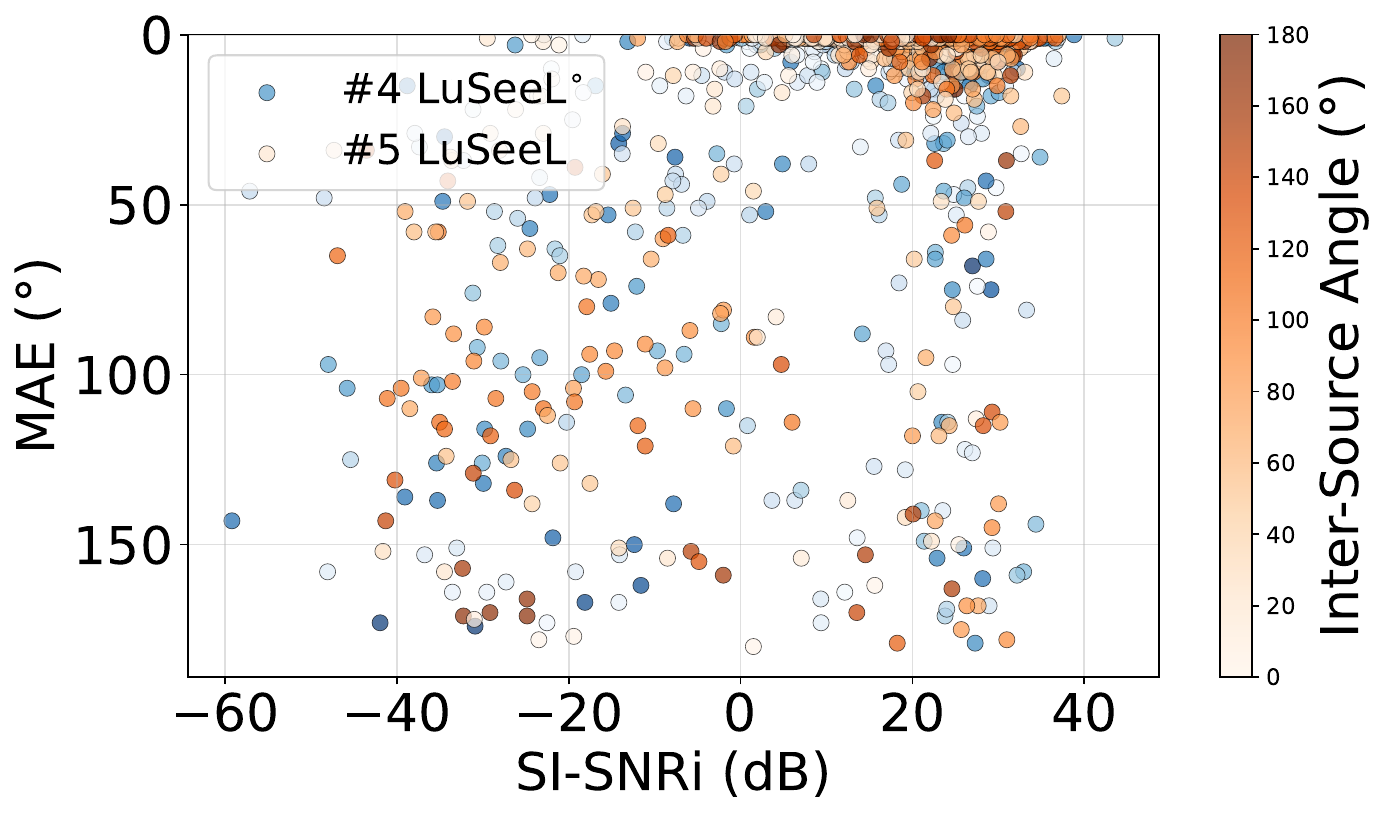}}
  \vspace*{-2mm}
  \caption{The 2-source MAE against SI-SNRi scatter plot.}\medskip
  \label{fig:scatter}
\end{minipage}
\end{figure*}

\subsection{Comparison with baselines}

We first analyze the 2-source mixture results. System 1 is a single-channel extraction baseline, achieving a poor SI-SNRi of 7.7 dB. In contrast, System 3, which uses binaural audio, improves extraction performance significantly to 17.6 dB. This highlights the importance of dual-channel inputs in separating audio sources, particularly when they originate from different angular directions.

System 5 is our proposed model that jointly performs sound extraction and localization. It outperforms System 3 by 2.7 dB in SI-SNRi and significantly surpasses System 2 (a dedicated localization network) in both accuracy and MAE. This demonstrates that joint training enhances performance on both tasks. Notably, for localization, the spatially aware embeddings learned during extraction provide stronger cues than handcrafted GCC-PHAT features. This is further supported by System 4, where removing the GCC-PHAT encoder results in minimal degradation in either extraction or localization performance, indicating that the extraction network implicitly learns effective spatial representations.

Similar trends are observed in the 3-source mixture results, except that System 9 outperforms System 10 by 0.4 dB in extraction performance (SI-SNRi), while localization accuracy remains better when using GCC-PHAT features (System 10). This indicates that GCC-PHAT still provides complementary spatial information, particularly beneficial in more complex acoustic scenes.

\subsection{Visualization}
In Fig.~\ref{fig:histogram_sisnr}, we visualize the extraction performance (SI-SNRi) of Systems 3 and 5 on 2-source mixtures. System 5 consistently outperforms System 3 across various inter-source angular separations, demonstrating the benefit of joint training with the localization task.

In Fig.~\ref{fig:histogram_mae}, we show the localization MAE of Systems 2 and 5 on the 2-source mixtures. The system 2 MAE increases with larger angular separation between sources, which we attribute to confusion errors, i.e., the model often predicts the direction of the interfering sound instead of the target. This suggests difficulty in aligning the text query with the correct spatial source via GCC-PHAT features. In contrast, System 5 achieves consistently low MAE, indicating that the audio extraction features and end-to-end joint learning help resolve such ambiguities.

In Fig.~\ref{fig:scatter}, we plot MAE against SI-SNRi for Systems 4 and 5, revealing error distributions. Most samples with high SI-SNRi (accurate extraction) have low MAE (precise localization), while low SI-SNRi samples tend to have higher MAE. However, many samples with bad MAE still achieve high SI-SNRi even when the extraction is good. This suggests that poor extraction degrades localization performance, but inaccurate localization has a weaker impact on extraction, indicating a degree of asymmetry in task dependency.

\section{Conclusion}
We presented a language-driven framework for joint binaural sound extraction and localization. Our results show that leveraging binaural input significantly improves extraction performance, and that joint training enhances both extraction and localization through shared spatial representations. While GCC-PHAT provides useful spatial cues, our model demonstrates that end-to-end learning can effectively align language queries with target sources, reducing reliance on handcrafted features. This work highlights the potential of multimodal, spatially aware models for intuitive auditory scene understanding.

\newpage

\bibliographystyle{IEEEbib}
\bibliography{IEEEabrv,Bibliography}

\end{document}